\documentclass[twocolumn,aps,prc,superscriptaddress]{revtex4}
\usepackage{amsmath,bm}
\usepackage{graphicx}
\usepackage{dcolumn}
\usepackage{multirow}
\begin{document}
\draft
\title{Scaling of Anisotropic Flows and Nuclear Equation of State in Intermediate Energy Heavy Ion
Collisions}

\thanks{Supported partially by  the
National Natural Science Foundation of China under Grant No
10535010 and 10610285, the Shanghai Development Foundation for
Science and Technology under Grant Numbers 06JC14082 and
05XD14021, and CAS project KJCX3.SYW.N2}

\author{Yan Ting-Zhi}
\affiliation{Shanghai Institute of Applied Physics, Chinese
Academy of Sciences, P. O. Box 800-204, Shanghai 201800, China}
\affiliation{Graduate School of the Chinese Academy of Sciences,
Beijing 100039, China}
\author{Ma Yu-Gang} \thanks{Corresponding author. Email: ygma@sinap.ac.cn}
  \affiliation{Shanghai Institute of Applied Physics, Chinese Academy of Sciences, P. O. Box 800-204,
Shanghai 201800, China}
\author{Cai Xiang-Zhou}
\affiliation{Shanghai Institute of Applied Physics, Chinese
Academy of Sciences, P. O. Box 800-204, Shanghai 201800, China}
\author{Fang De-Qing}
\affiliation{Shanghai Institute of Applied Physics, Chinese
Academy of Sciences, P. O. Box 800-204, Shanghai 201800, China}
\author{Guo Wei}
\affiliation{Shanghai Institute of Applied Physics, Chinese
Academy of Sciences, P. O. Box 800-204, Shanghai 201800, China}
\affiliation{Graduate School of the Chinese Academy of Sciences,
Beijing 100039, China}
\author{Ma Chun-Wang}
\affiliation{Shanghai Institute of Applied Physics, Chinese
Academy of Sciences, P. O. Box 800-204, Shanghai 201800, China}
\affiliation{Graduate School of the Chinese Academy of Sciences,
Beijing 100039, China}
\author{Shen Wen-Qing}
  \affiliation{Shanghai Institute of Applied Physics, Chinese Academy of Sciences, P. O. Box 800-204,
      Shanghai 201800, China}
 \author{Tian Wen-Dong}
 \affiliation{Shanghai Institute of Applied Physics, Chinese Academy of Sciences, P. O. Box 800-204,
   Shanghai 201800, China}
\author{Wang Kun}
\affiliation{Shanghai Institute of Applied Physics, Chinese
Academy of Sciences, P. O. Box 800-204, Shanghai 201800, China}

\date{\today}

\begin{abstract}
Elliptic flow ($v_2$)  and hexadecupole flow ($v_4$) of light
clusters have been studied in details for 25 MeV/nucleon $^{86}$Kr
+ $^{124}$Sn at large impact parameters by Quantum Molecular
Dynamics model with different potential parameters. Four parameter
sets which include soft or hard equation of state (EOS)
with/without symmetry energy term are used. Both number-of-nucleon
($A$) scaling of the elliptic flow versus transverse momentum
($p_t$) and the scaling of $v_4/A^{2}$ versus $(p_t/A)^2$ have
been demonstrated for the light clusters in all above calculation
conditions. It was also found that the ratio of $v_4/{v_2}^2$
keeps a constant of 1/2 which is independent of $p_t$ for all the
light fragments. By comparisons among different combinations of
EOS and symmetry potential term, the results show that the above
scaling behaviors are solid which do not depend the details of
potential, while the strength of flows is sensitive to EOS and
symmetry potential term.\vspace{1.cm}


\textbf{PACC}: 2410, 2570, 2587
\end{abstract}


\maketitle


Anisotropic flows are very useful to explore heavy-ion collision
dynamics  since it results from the transition of the original
space-time asymmetry into a momentum space anisotropy for the
non-central collision \cite{J. Ollitrault,Y. G. Ma4, W. Q. Shen,H.
Sorge,P. Danielewicz,D. Teaney,P. F. Kolb,Y. Zheng,Li,D.
Perslam,J. Lukasik,J.Adams,J. H. Chen,Ma06}. Different mechanisms
will contribute the final momentum anisotropy, i.e. flow.  Many
studies of the dependence of the directed flow ($v_1$) and  the
elliptic flow ($v_2$) on beam energies, mass number, isospin and
impact parameter have been carried out and much interesting
physics has been demonstrated on the properties and origin of the
collective motion in both nucleonic or partonic levels. Very
recently, we carried out a Quantum Molecular Dynamics model
calculation with hard equation of state  and symmetry energy term
and found that there is a nucleon-number scaling for the elliptic
flow of light particles up to the mass number A = 4
\cite{Yan_PLB}.  In this work, we shall present more details for
the nucleon number dependence of the anisotropic flows $v_2$ and
$v_4$  for $^{86}$Kr + $^{124}$Sn collisions at 25 MeV/nucleon and
large impact parameters ($b = 7-10 fm$) with different EOS and
symmetry energy interaction. The scaling behaviors look robust
since they do not depend the parameters used in the model, and the
sensitivities of EOS and symmetry potential for the $v_2$ and
$v_4$ are discussed.

Anisotropic flow is defined as the different $n$-th harmonic
coefficient $v_n$ of the Fourier expansion for the particle
invariant azimuthal distribution
\begin{equation}
\frac{dN}{d\phi}\propto{1+2\sum^\infty_{n=1}{v_n\cos(n\phi)}},
\end{equation}
where $\phi$ is the azimuthal angle between the transverse
momentum of the particle and the reaction plane. Note that the
z-axis is defined as the direction along the beam and the impact
parameter axis is labelled as x-axis. The first harmonic
coefficient $v_1$ represents directed flow, $v_1 = \langle cos\phi
\rangle = \langle \frac{p_x}{p_t} \rangle$, where $p_t =
\sqrt{p^2_x+p^2_y}$ is transverse momentum. While the $v_2$ which
measures the eccentricity of the particle distribution in the
momentum space represents elliptic flow,
\begin{equation}
v_2 = \langle cos(2\phi) \rangle = \langle
\frac{p^2_x-p^2_y}{p^2_t} \rangle,
\end{equation}
and $v_4$ represents the 4-th momentum anisotropy, namely
hexadecupole flow:
\begin{equation}
v_{4} =\left\langle \frac{p_{x}^{4}-6p_{x}^{2}p_{y}^{2}+p_{y}^{4}}{%
p_{t}^{4}}\right\rangle . \label{v4}
\end{equation}

The intermediate energy heavy-ion collision dynamics is complex
since both mean field and nucleon-nucleon collisions are playing
the competition roles. Furthermore, the isospin dependent role
should be also incorporated for asymmetric reaction systems.
Isospin dependent Quantum Molecular Dynamics model (IDQMD) has
been affiliated with isospin degrees of freedom with mean field
and nucleon-nucleon collision \cite{J. Aichelin,Y. G.
Ma1,ZhangFS,J.Y. Liu,H. Y. Zhang,Y. B. Wei,Y. G. Ma2,Y. G. Ma3}.
The IDQMD model can explicitly represent the many body state of
the system and principally contain correlation effects to all
orders and all fluctuations, and can describe the time evolution
of the colliding system well. When the spatial distance ($\Delta
r$) is closer than 3.5 fm and the momentum difference ($\Delta p$)
is smaller than 300 MeV/c between two nucleons, two nucleons can
coalesce into a cluster \cite{J. Aichelin}. With this simple
coalescence mechanism which has been extensively applied in
transport theory, different size clusters can be recognized.

In the model the nuclear mean-field potential is parameterized as
\begin{equation}
U(\rho,\tau_{z}) = \alpha(\frac{\rho}{\rho_{0}}) +
\beta(\frac{\rho}{\rho_{0}})^{\gamma} +
\frac{1}{2}(1-\tau_{z})V_{c} \nonumber
\end{equation}
\begin{equation}
+ C_{sym} \frac{(\rho_{n} - \rho_{p})}{\rho_{0}}\tau_{z} + U^{Yuk}
\end{equation}
where $\rho_0$ is the normal nuclear matter density
($0.16fm^{-3}$), $\rho_n$, $\rho_p$ and $\rho$  are the  neutron,
proton and total densities, respectively. $\tau_z$ is $z$-th
component of the isospin degree of freedom, which equals 1 or -1
for neutrons or protons, respectively. The coefficients $\alpha$,
$\beta$ and $\gamma$ are parameters for nuclear equation of state.
$C_{sym}$ is the symmetry energy strength due to the density
difference of neutrons and protons in nuclear medium, which is
important for asymmetry nuclear matter \cite{Ma-acta,C. Zhong,G.
C. Yong} (here $C_{sym} = 32$ MeV is used to consider symmetry
energy effect or isospin-dependent potential, and $C_{sym} = 0$
for no symmetry energy effect or isospin-independent potential).
$V_c$ is the Coulomb potential and $U^{Yuk}$ is Yukawa (surface)
potential. In this work, we take $\alpha $ = 124 MeV, $\beta$ =
70.5 MeV and $\gamma$ = 2 which corresponds to the so-called hard
EOS with an incompressibility of $K$ = 380 MeV, and $\alpha$ =
-356 MeV, $\beta$ = 303 MeV and $\gamma$ = 7/6 which corresponds
to the so-called soft EOS with an incompressibility of $K$ = 200
MeV. In the present study,  four combinations with different
potential parameters, i.e. parameters of hard or soft EOS  with or
without
 symmetry energy effect (i.e. $C_{sym} = 32$ or 0 MeV),
 for the collision system of $^{86}$Kr + $^{124}$Sn
at 25 MeV/nucleon with impact parameter from 7 fm to 10 fm were
carried out. The physics results were extracted at the time of 200
fm/c when the system has been in the freeze-out stage.

The Fig.1 (a), (b), (e) and (f) shows transverse momentum
dependence of elliptic flows for mid-rapidity light fragments in
four different calculation conditions: (a)  for soft EOS with
 symmetry potential ($soft\_iso$);
 (b) for hard EOS with symmetry  potential ($hard\_iso$);
 (e) for soft EOS without symmetry  potential
 ($soft\_niso$) and
 (f) for hard EOS without  symmetry potential ($hard\_niso$).
In all cases, elliptic flow is positive and it increases with the
increasing $p_t$, which is apparently similar to RHIC's results
\cite{J.Adams,J. H. Chen}. Of course, the mechanism is very
different. In intermediate energy domain, collective rotation is
one of the main mechanisms to induce the positive elliptic flow
\cite{J. P. Sullivan,W. Q. Shen,Y. G. Ma4,R. Lacey, Z. Y. He}.
However, at RHIC energies it is the strong pressure which is built
in early initial almond anisotropy of the geometrical overlap zone
between both colliding nuclei that drives the positive elliptic
flow \cite{J.Adams}. The
 corresponding nucleon-number scaled elliptic flows are plotted in Fig.1 (c), (d), (g) and (h) as a
function of transverse momentum per nucleon. From these panels, it
seems that the number of nucleon scaling for elliptic flow exists
for light fragments at low $p_t/A$ ($p_t/A<0.2GeV/c$). This
behavior is apparently similar to the number of constituent quarks
scaling of elliptic flow versus transverse momentum per
constituent quark ($p_t/n$) for different mesons and baryons which
was observed at RHIC \cite{J.Adams}. Since all calculations show
the similar scaling behavior, this scaling behavior is robust, and
it is independent of the details of EOS and symmetry potential.
\begin{figure}
\vspace{-0.1truein}
\includegraphics[scale=0.7]{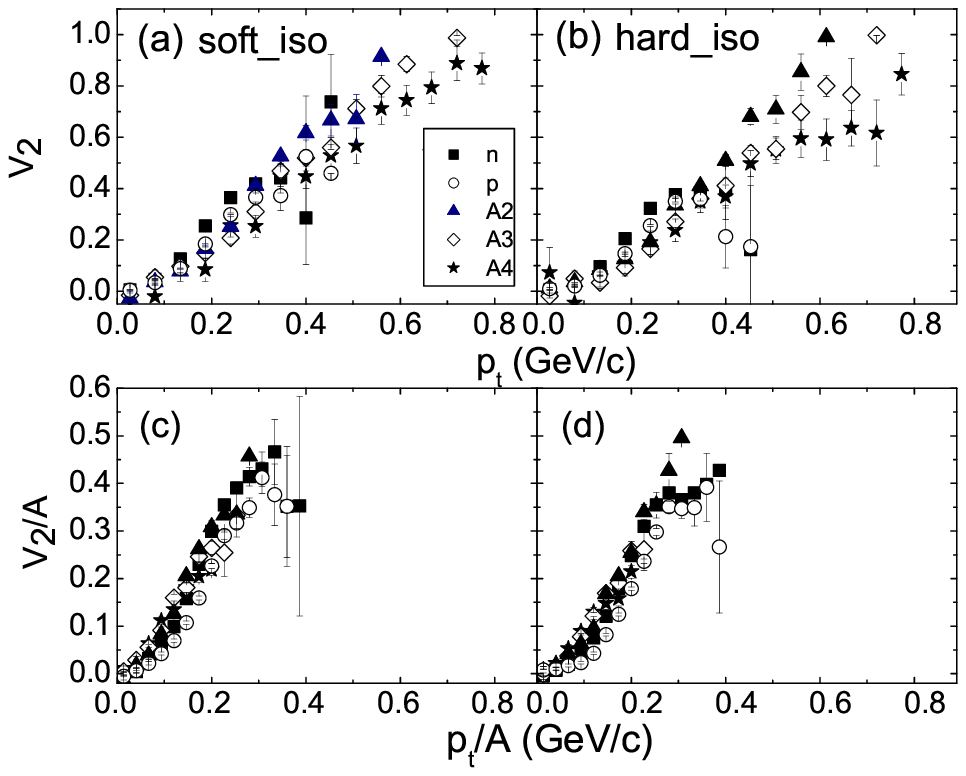}
\vspace{-0.15truein}
\includegraphics[scale=0.7]{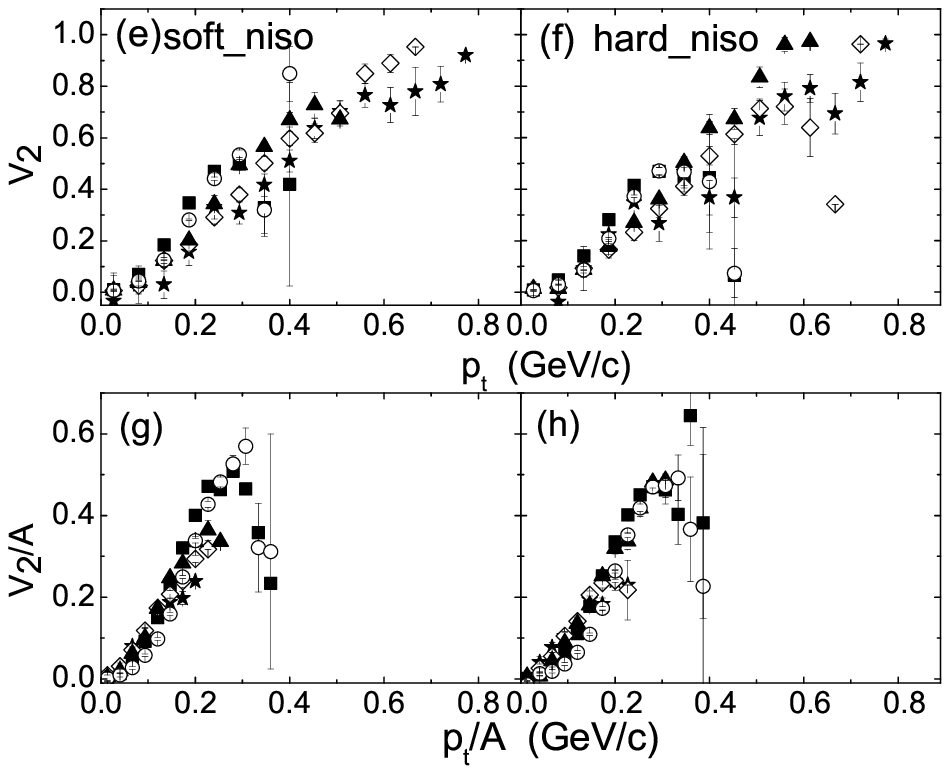}
 \caption{\footnotesize(a), (b), (e) and (f): Elliptic flow as a function of transverse momentum
 ($p_t$) for the
 simulation with different parameters of EOS with or without symmetry energy term.
 (a)  for soft EOS with
 symmetry potential ($soft\_iso$);
 (b) for hard EOS with symmetry  potential ($hard\_iso$);
 (e) for soft EOS without symmetry  potential
 ($soft\_niso$); and
 (f) for hard EOS without  symmetry potential ($hard\_niso$).
Squares represent for neutrons, circles for protons, triangles for
fragments of A=2, diamonds for A=3 and stars for A=4.
 Fig.(c), (d), (g) and (h) presents nucleon-number normalized
 elliptic flow as a function of
transverse momentum per nucleon corresponding to the case of (a),
(b), (e) and (f), respectively.  }
\end{figure}

\begin{figure}
\vspace{-0.9truein}
\includegraphics[scale=0.45]{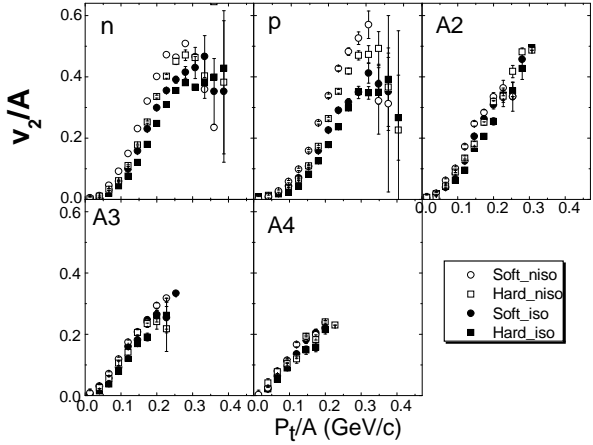}
\vspace{-2.1truein}
 \caption{\footnotesize Comparisons of the values of $v_2/A$ versus $p_t/A$ in the different
 simulation conditions. The meanings of
 the symbols are depicted in right bottom corner.}
\end{figure}

To quantitatively look the difference of the flows in different
calculation conditions, we compare the values of $v_2/A$ for  the
four simulation conditions (see Fig.2). The figures show that the
difference between different simulations is big for neutrons and
protons but a little small for the fragments of $A=2$, $A=3$ and
$A=4$. The reason is that the emitted protons and neutrons can
feel the role of mean field (EOS) directly, while the light
fragments have weak sensitivity since they are indirected products
 by the coalescence mechanism in the present model.
Approximately  at the same $p_t/A$, the elliptic flow is larger
for soft EOS than the one for hard EOS, and it is larger for EOS
with symmetry potential than the case  without symmetry potential.
Considering that the symmetry potential is basically positive for
the studied reaction system (more neutrons than protons), symmetry
potential will make the whole EOS stiffer. In this case, we can
say that the stiffer the EOS, the smaller the flow. In other
words, we can say the strength of elliptic flow per nucleon is
sensitive to the EOS and symmetry potential.

\begin{figure}
\vspace{-0.1truein}
\includegraphics[scale=0.7]{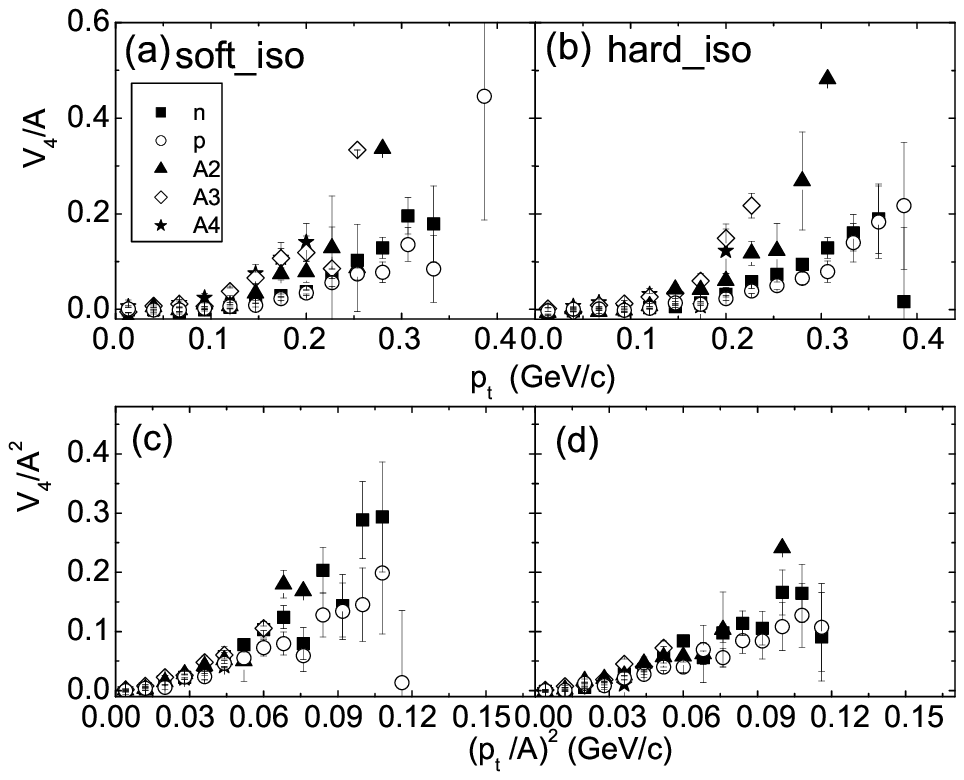}
\includegraphics[scale=0.7]{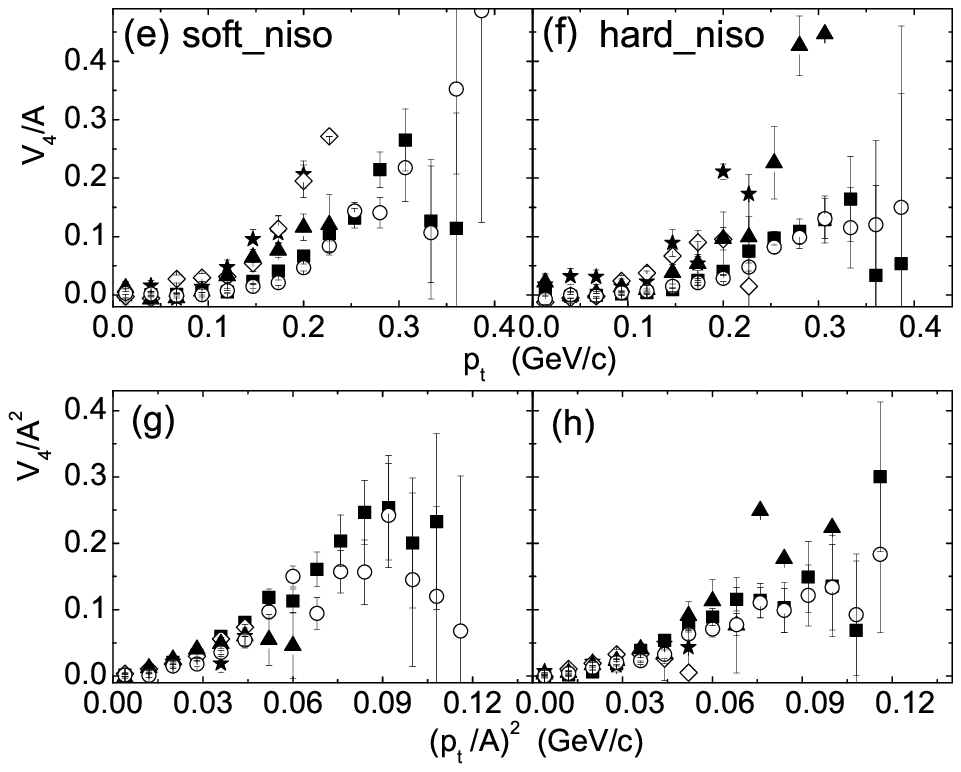}
 \caption{\footnotesize Same as Fig.1 but for $v_4/A$ versus $p_t$ [(a), (b), (e) and (f)] and $v_4/A^2$ versus $(p_t/A)^2$ [
 (c), (d), (g) and (h)].}
\end{figure}

So far, there is rare studies  about higher order flows, such as
$v_4$, experimentally and theoretically in this energy domain.
Here we try to explore the behavior of $v_4$. First we draw
$v_4/A$ as a function of $p_t/A$ mimicing  the behavior of
elliptic flow (see (a), (b), (e) and (f) of Fig.3) for four
different calculation conditions. It shows that $v_4/A$ is
positive and increases with $p_t/A$, but there seems  no simple
scaling behavior as $v_2$ shows. Considering that RHIC
experimental data have demonstrated that a scaling relation among
hadron anisotropic flows holds, i.e., $v_n(p_t)\sim
v^{n/2}_2(p_t)$ \cite{J. Adams2}, we plot $v_4/A^2$ as a function
of $(p_t/A)^2$
 in Fig. 3 (c), (d), (g) and (h) for the
corresponding calculation conditions of Fig.3(a), (b), (e) and
(f). Now the points of different light fragments nearly merge
together at low $(p_t/A)^2$, which means a certain of scaling law
holds between two variables. All the calculation cases show that
there is the  scaling behavior for $v_4/A^2$ versus $(p_t/A)^2$,
and this behavior is robust regardless the parameters which we
used for EOS.

\begin{figure}
 \vspace{-0.1truein}
\includegraphics[scale=0.7]{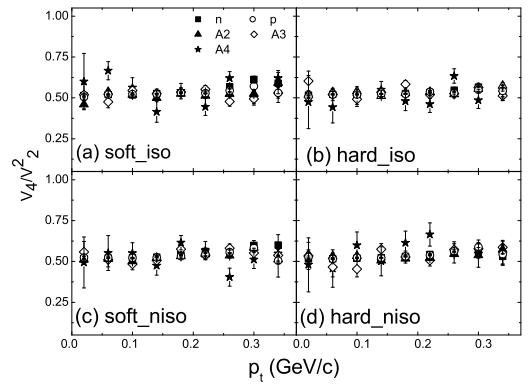}
 \caption{\footnotesize The ratios of $v_4/v^2_2$ for
neutrons (squares), proton (circles), the fragments of $A=2$
(triangles), $A=3$ (diamonds) and  $A=4$ (stars) versus $p_t$ for
the four simulations in different calculation conditions.}
\end{figure}

Since the above scaling behavior  assumes $v_n(p_t)\sim
v^{n/2}_2(p_t)$, so we plot $v_4/v^2_2$ as a function of $p_t$ in
Fig. 4 for the four simulations. The figures show that the ratios
of $v_4/v^2_2$ for different fragments up to $A=4$  are about a
constant of $1/2$ in all simulation cases. Because $v_2/A$ can be
scaled with $p_t/A$,  $v_4/A^2$ should scale versus $(p_t/A)^2$,
which is exactly what we see in Fig. 4. One point is worth to be
mentioned comparing to the RHIC studies where the data shows
$v_4/v^2_2\sim 1.2$ \cite{ J. Adams2}, $v_4/v^2_2\sim 1/2$  for
the light nuclear fragments in this nucleonic level coalescence
mechanism rather than the value of 3/4 for mesons or 2/3 for
baryons in quark coalescence model \cite{kolb}.  Coincidentally,
the predicted value of the ratio of $v_4/v_2^2$ for hadrons is
also 1/2 if the matter produced in ultra-relativistic heavy ion
collisions reaches to thermal equilibrium and its subsequent
evolution follows the laws of ideal fluid dynamics \cite{Bro}. It
is interesting to note the same ratio was predicted in two
different models at very different energies, which is of course
worth to be further investigated in near future. One possible
interpretation is that the big nucleon-nucleon cross sections in
low energy HIC make the system to reach thermal equilibrium and
may induce the fluid-like behavior of nuclear medium before the
light fragments are coalesced by nucleons. In this case, the value
of $v_4/v_2^2$ of light fragments could be  $\sim$ 1/2 as
Ref.\cite{Bro} shows.

The values of $v_4/A$ versus $p_t/A$ with  different simulation
parameters are also presented for light fragments, see Fig.5. The
figures are similar to those in Fig.2, and  the effects of EOS and
symmetry potential on $v_4/A$ are also similar to their effects on
$v_2$. However, comparing with the $v_2$'s sensitivity to the EOS
and symmetry potential, $v_4/A$ is not so salient.

\begin{figure}
 \vspace{-0.9truein}
\includegraphics[scale=0.45]{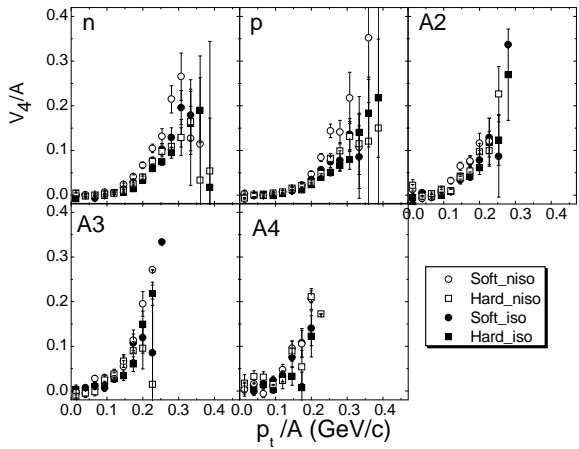}
\vspace{-2.0truein}
 \caption{\footnotesize Comparison of the values of $v_4/A$ versus $p_t/A$. The meanings of
 the symbols are depicted in right bottom corner.}
\end{figure}

To summarize, we investigated the behavior of anisotropic flows as
a function of  transverse momentum for light fragments for the
simulations of 25 MeV/nucleon $^{86}Kr$+$^{124}Sn$ collisions in
peripheral collisions by IDQMD model in the potential parameters
of hard or soft EOS  with or without symmetry energy term. It was
found that for all the four type simulations $v_2$ and $v_4$  of
light fragments are positive and increase with $p_t/A$. When we
plot $v_2$ per nucleon ($v_2/A$) versus $p_t/A$ for all light
particles, all curves collapse onto the same curve. Similarly, the
values of $v_4/A^2$ merge together as a function of $(p_t/A)^2$
for all light particles. Furthermore, it was found that $v_4$ can
be well scaled by $v^2_2$, and the value of $v_4/v^2_2\sim 1/2$
which does not depend on transverse momentum. The above scaling
behaviors can be seen as an outcome of the nucleonic coalescence,
and it illustrates that the number-of-nucleon scaling for elliptic
flow exists in intermediate energy heavy ion collision. In
addition, the values of $v_2/A$ and $v_4/A$ were compared in
different simulation conditions, and it was shown that the values
of the $v_2$  are sensitive to the EOS and symmetry potential,
especially for neutrons and protons.

{}
\end{document}